\documentclass{article}
\usepackage[a4paper, total={6in, 8in}]{geometry}
\usepackage[utf8]{inputenc}
\usepackage{amsmath}
\usepackage{amssymb}
\usepackage{caption}
\usepackage{subcaption}
\usepackage{float}
\usepackage{graphicx}
\usepackage{cancel}
\usepackage{xcolor}
\usepackage{natbib}

\title{Does Gaia Play Dice? : Simple Models of non-Darwinian Selection}

\author{Rudy Arthur,$^{1\ast}$ Arwen Nicholson,$^{2}$\\
{$^{1}$University of Exeter, Department of Computer Science}\\
{Stocker Rd, Exeter EX4 4PY}\\
{$^{2}$University of Exeter, Department of Physics and Astronomy}\\{Stocker Rd, Exeter EX4 4PY}\\
{$^\ast$To whom correspondence should be addressed;}\\
{E-mail:  r.arthur@exeter.ac.uk}
}

\begin{document} 

\maketitle

\begin{abstract}
In this paper we introduce some simple models, based on rolling dice, to explore mechanisms proposed to explain planetary habitability. The idea is to study these selection mechanisms in an analytically tractable setting, isolating their consequences from other details which can confound or obscure their effect in more realistic models. We find that while the observable of interest, the face value shown on the die, `improves' over time in all models, for two of the more popular ideas, Selection by Survival and Sequential Selection, this is down to sampling effects. A modified version of Sequential Selection, Sequential Selection with Memory, implies a statistical tendency for systems to improve over time. We discuss the implications of this and its relationship to the ideas of the `Inhabitance Paradox' and the `Gaian bottleneck'.
\end{abstract}

\section{Introduction}

Recent discussion about the persistence of life over geological time spans has brought to the fore various selection principles \citep{bourrat2014survivors, ford2014natural, toman2017stability, lenton2018selection, arthur2022selection}. Distinct from Darwinian selection, which involves reproduction of individuals in some environment, these selection principles consider a (real or imagined) population of non-reproducing, potentially immortal entities. Such entities are suggested to be species/lineages \citep{ford2014natural}, biogeochemical cycles \citep{doolittle2017darwinizing} or, as we will consider here, the entire life-earth coupled system \citep{lenton2018selection}, namely, Gaia \citep{lovelock1974atmospheric}. 

While selection principles have mostly been discussed with reference to Earth history \citep{lenton2018selection, boyle2022evolution}, the recent launch of the James Webb Space Telescope \citep{Snellen:2021, Quanz:2021} brings questions of astrobiology into focus. Until it is possible to experimentally investigate a large catalogue of inhabited planets \citep{lenardic2016solar}, understanding general principles behind planetary inhabitance is a way to provide working hypotheses that can explain both how `lucky' the Earth is to be inhabited  and what we might expect on planets orbiting other stars.
 
The most discussed of these selection principles is called `Selection by Survival' (SBS) in \cite{lenton2018selection} though the idea has many names, see \citep{toman2017stability}. The essential idea is that a population where the entities have different rates of survival will be `purified' so that, in the long run, surviving entities must have properties conducive to survival. Several works e.g. \citep{bourrat2014survivors, ford2014natural, bouchard2014ecosystem, doolittle2017darwinizing},  attempt to disentangle Darwinian selection from SBS. These works emphasise the importance of SBS for higher order phenomena like ecosystems, biogeochemical cycles and Gaia. They argue that, in systems of hereditary replicators, Darwinian selection is more powerful but for entities without heredity and reproduction SBS will operate to favour certain macroscopic features. Specific examples suggested by \cite{toman2017stability} are sexual reproduction and macroevolutionary freezing.

\cite{lenton2018selection} defines another, related, selection principle called Sequential Selection (SS) \citep{betts2008second}. This is a similar idea to SBS, but motivated by the frequent upheavals in the history of life on Earth and meant to account for life's apparent stabilising effect on Earth's habitability. \cite{lenton2018selection} propose a simple algorithm - evolutionary innovations have a stabilizing or destabilizing effect on the environment. If they have a destabilizing effect,  habitability is reduced, eventually eliminating the destabilizing innovation. In this way destabilizing effects are eliminated by `near fatal resets' while stabilizing innovations persist and accumulate.

In \cite{arthur2017entropic,arthur2022selection} we argue for a refinement of this algorithm, emphasising that the resets are \textbf{near} fatal so the evolutionary innovations developed during the previous stable period are not completely lost. The algorithm of \cite{lenton2018selection} applies: destabilizing innovations lead to resets which greatly reduce species abundance but have a lesser effect on species diversity. The life-earth system which arises after the reset is selected from a larger `pool', which has the potential to generate larger, more stable ecosystems. Higher species and functional diversity give Gaia more tools to generate stability. The process is completely blind, so unstable states can also be selected, however, by definition, these are short lived and eventually a long-lived stable state will arise. During stable periods species diversity can increase again, leading to a kind of ratcheting effect. To emphasise this cumulative process, in contrast to the sequential selection algorithm of \cite{lenton2018selection}, we call this `Sequential Selection with Memory' (SSM).

A variety of abstract models of varying complexity have been proposed to explore these selection principles e.g. \citep{bourrat2014survivors,nicholson2018alternative,lenton2021survival,arthur2022selection,boyle2022evolution}.
Though these models have great value, it can sometimes be unclear what is a feature of a specific model and what is a generic feature. Here we propose using extremely simple probability models to study  selection principles. The aim is to strip out as much complexity as possible to understand the core of these principles and their consequences. A very loose analogy would be trying to understand the approximately Gaussian distribution of human height. A detailed model accounting for genetic and environmental causes could, with great difficultly, be formulated into a mechanistic model of height. However, a much simpler, and in many ways more satisfactory, explanation is that height is a sum of many independent effects and a Gaussian distribution is the expected outcome for heights in a population.  
 
A closer analogue is the classic Wright-Fisher model of population genetics \citep{wright1931evolution}.  The Wright-Fisher model takes $N$ diploid individuals and tracks allele frequency through a sequence of non-overlapping generations. The model neglects many important details: mutation, sexes, population size and (Darwinian) selection among others. Nonetheless, the Wright-Fisher model \emph{because of} its simplicity and tractability, provides a tool for studying more complex evolutionary forces and has been useful to understand subjects in population genetics like neutral theory \citep{kimura1983neutral} and the coalescent \citep{kingman1982coalescent}. The aim of this paper is to investigate the selection principles discussed above in a model as tractable as Wright-Fisher, so as to provide some clarity on exactly what they imply.

\section{Introducing the Models}

Consider an $M$ sided die with the rule that, once rolled, whatever number is showing on the top face gives the number of steps to wait before rolling again or finishing the game. For $r \geq 1$ dice we roll each one independently to get $x_1, x_2, \ldots, x_r$, and take the highest face value: $\max(x_1, x_2, \ldots, x_r)$. Based on these rules, consider the following dice games: 
\begin{enumerate}
\item \textbf{Selection By Survival(SBS):} Roll $N$ (where $N$ is a very large number) independent dice, once each.
\item \textbf{Sequential Selection (SS):} Roll one die repeatedly for $T$ time steps.
\item \textbf{Sequential Selection with Memory (SSM):} 
\begin{enumerate}
    \item Starting with $r=1$, roll $r$ dice repeatedly for $T$ time steps. Add a new die every time the top face shows the maximum value, $M$.
    \item Starting with $M=1$, roll an $M$ sided die repeatedly for $T$ times steps. Every time the top face shows the maximum value $M$, increase $M$ by 1.
    \end{enumerate}
\end{enumerate}
The SSM games are reminiscent of the Polya Urn model, though have not been studied before to our knowledge. The quantity of interest will be the expected face value at time $t$. The names chosen are based on the discussion in the Introduction and follow the conventions of \cite{lenton2018selection}. Our version of Selection By Survival is much simpler than the (mostly verbal) models proposed by others e.g. \cite{bouchard2014ecosystem} and most closely follows the graphical model from \cite{ford2014natural}. 

As a rough mapping to reality - a die represents an inhabited `planet'. Each roll is a period of stability for the planet's biosphere. The face value represents something akin to the `fitness' of the biosphere on that planet, i.e. how long it will persist. If we observe an inhabited planet at a random point in its history we may see a biosphere with properties conducive to long term stability (high face value) or only short term stability (low face value). The question of interest for astrobiology is, if we were to survey a large catalogue of inhabited planets, what is the expected `fitness'? For Earth history (or for the history of any inhabited planet) the equivalent question is, if we were to observe a planet at a random point in time, what should we expect about the habitability properties of the planet at that time? 

Note that we are being deliberately coarse in our mapping and our use of terms like `inhabited' and `stable'. More detailed models should make these terms precise (e.g. see \cite{Arthur:2023} for a discussion of habitability and stability in a model of a coupled population and environment). To avoid confusion - real planets are not like dice! However, understanding the consequences of selection principles acting on simple systems will give us some insight into what we might expect in more complex models, or real planetary history. To keep the presentation clear the key results are stated in the main paper, their derivation is given in the Appendix.

\section{Selection by Survival}

\begin{figure}[H]
     \centering
         \includegraphics[width=\textwidth]{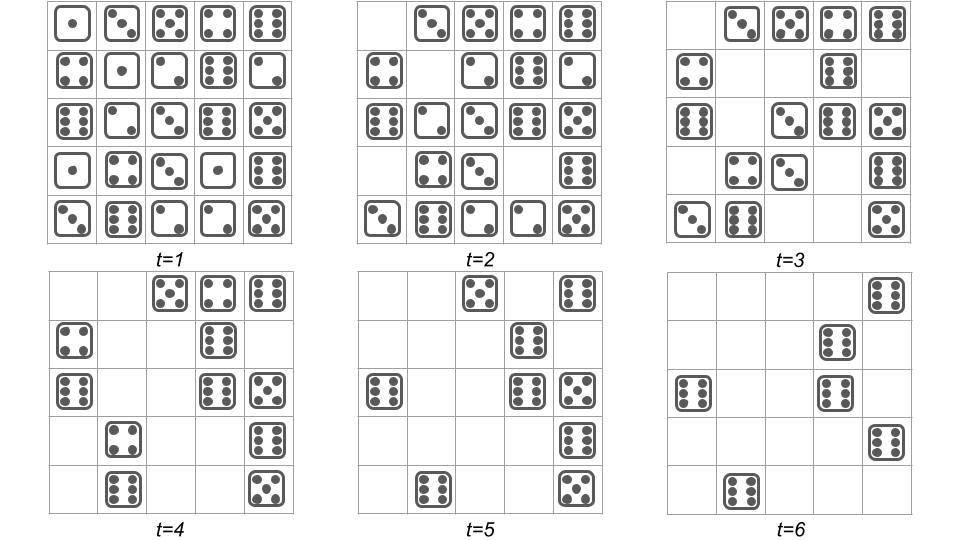}
        \caption{One possible unfolding of the SBS game with $N=25$ and $M=6$. At $t=1$ we have our initial ensemble, at $t=2$ we have removed all the 1s, at $t=3$ we remove all the 2s etc.}
        \label{fig:sbs_example}
\end{figure} 

\begin{figure}[H]
         \centering
         \includegraphics[width=\textwidth]{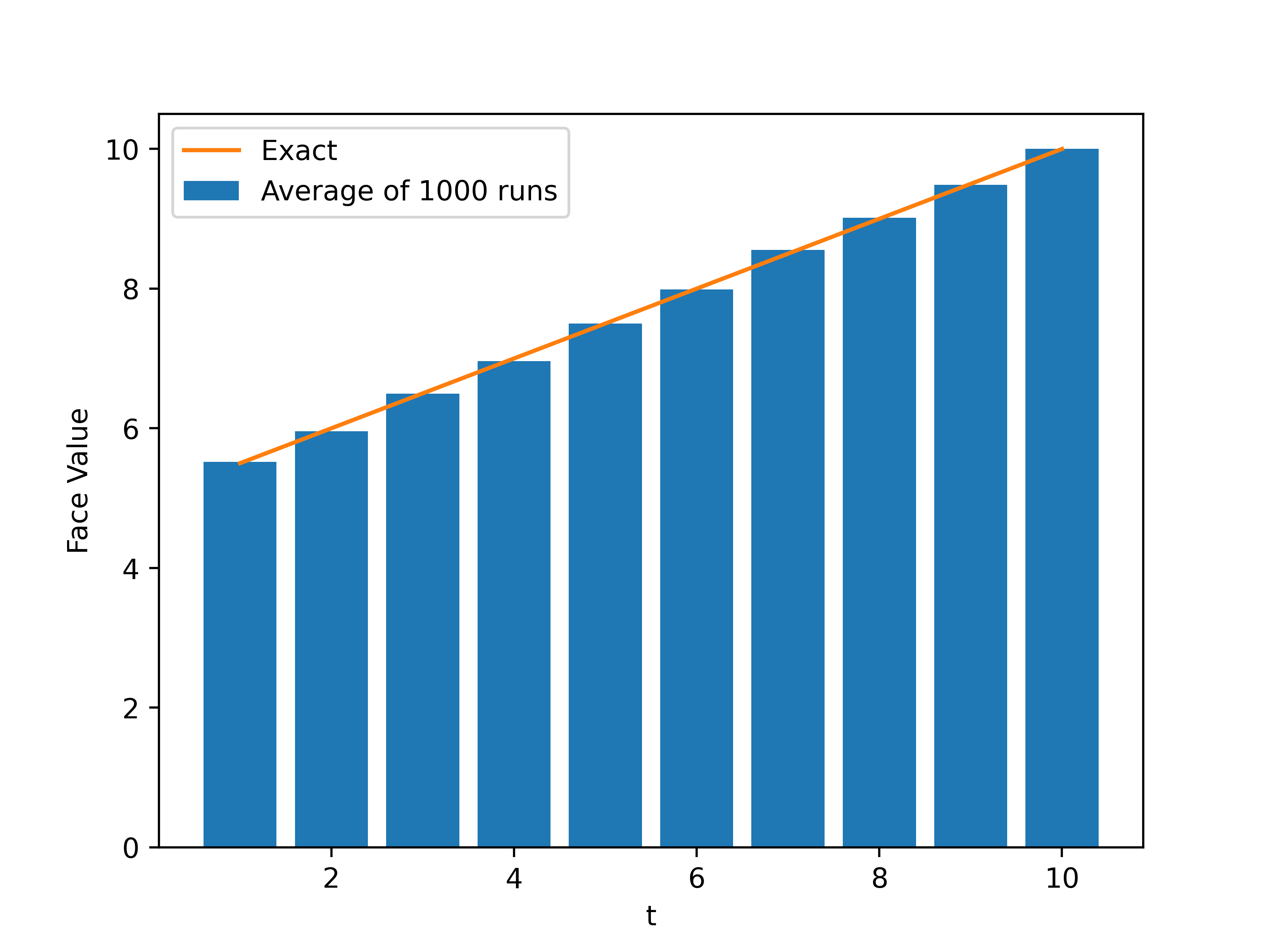}
         \caption{Average face value in the SBS game as a function of $t$ for an $M=10$ sided dice over $N=1000$ dice rolls.}
         \label{fig:sbs_run}
\end{figure}

Figure \ref{fig:sbs_example} shows one realisation of the SBS game. The expected face value at time $t$ is \begin{equation}\label{eqn:sbseqn}
E[f] (t) = \frac{M+t}{2}
\end{equation}
Figure \ref{fig:sbs_run} shows the result of $N=1000$ simulations of the game compared to equation \ref{eqn:sbseqn}. $N=1000$ is chosen in this and subsequent games to show variance arising from a finite sample while still being accurate enough to show any systematic deviation from the analytic solution. In terms of `planets' this model is simply stating the (obvious) fact that planets which survive have properties (high face value) which allow them to survive. Therefore, a catalogue of inhabited planets will necessarily contain planets with properties conducive to maintaining life, without the need for any additional mechanism. 

Here, all planets seeded with life at the same time. More complex models could be devised (say a constant rate of habitable planet generation) to study how the generation rate interacts with this simple selection mechanism. For this paper, SBS represents a basic null model - older inhabited planets must have features which have enabled them to remain inhabited. The growth in fitness of the surviving planets is a sampling artefact. SBS has little to say about our own solar system. The SBS approach simply says that we are observing at a late time, when the attempts at life on, say, Mars and Venus, have already ended in failure. Life on Earth persists because of some unspecified persistence enhancing feature or dumb luck. Thus SBS is basically an anthropic principle and seems to have little to say for exoplanet surveys.

\section{Sequential Selection}

\begin{figure}[H]
         \centering
         \includegraphics[trim={0cm 8cm 0cm 8cm},clip,width=\textwidth]{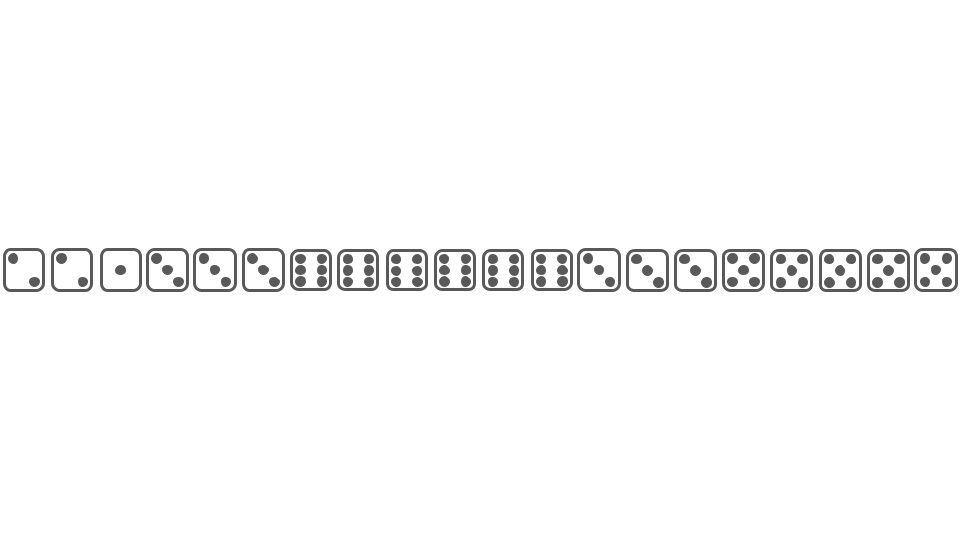}
         \caption{One possible unfolding of the SS game with $T=20$ and $M=6$. At $t=1$ we roll 2 which shows for 2 steps, we roll 1 which shows for 1 step, then 3 for 3 steps etc. The game is played a large number, $N$, of times as in Figure \ref{fig:sbs_example} and we are interested in the average behaviour. }
         \label{fig:ss_example}
\end{figure}

The Earth has experienced numerous mass extinctions, had very different planetary regulation mechanisms, atmospheric composition, levels of volcanic activity and life has persisted the entire time \citep{lenton2013revolutions}. We seek to model these sequential resets with another simple game: repeatedly rolling a single die. 
Figure \ref{fig:ss_example} is one realisation of the sequential selection game. The expected face value of the die at $t$ is 
\begin{equation}\label{eqn:ss_eqn}
E[f] (t)= \frac{2M+1}{3}
\end{equation}

\begin{figure}[H]
         \centering
         \includegraphics[width=\textwidth]{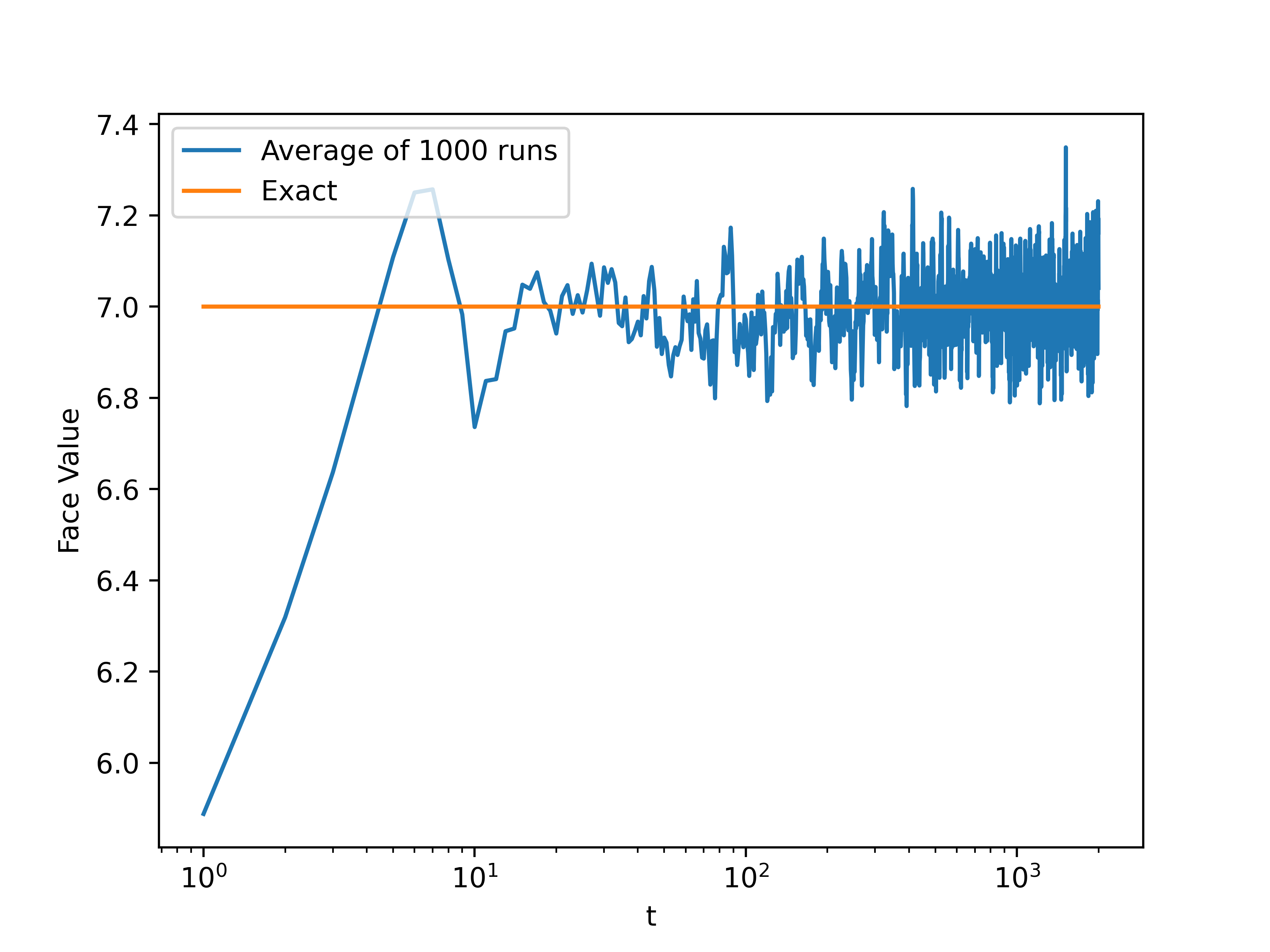}
         \caption{Average face value in the SS game as a function of $t$ for an $M=10$ sided dice, averaged over $N=1000$ independent instances. The expected value of a single $10$ sided dice roll is $5.5$ which is less than the expected face value in the sequential sampling game, $7$. Note the logged x-axis.}
         \label{fig:ss_run}
\end{figure}
Figure \ref{fig:ss_run} shows the results of simulations of the game compared to equation \ref{eqn:ss_eqn}. The results here imply that when observing a `planet' at a random time, it is more likely to be in a state with stability enhancing properties (high face value). Again this reflects the obvious fact that if we depict Earth's history as a time-line and pick a random point on the line, we are more likely to pick a point in a long, stable period. In particular, our present time is most likely to be a stable period, without the need for any additional mechanism. Like with SBS, Earth's current stability is simply an observer effect.

One thing missing from this game is the possibility of total extinction. We could implement an additional rule, say when we roll a 1, stop the game. This would give a model where SBS and SS are both operating simultaneously. For simplicity we don't implement this, so as not to mix the mechanisms. More complex models \citep{arthur2022selection, Arthur:2023} do have the possibility of early stopping and come to very similar conclusions as above. 

\section{Sequential Selection with Memory}

The continued inhabitance of Earth and the fact that biodiversity has increased over time motivates the final games. Each reset does not start from scratch, but builds on evolutionary and ecological innovations that came before. We propose 2 models with an extremely simple `memory'. This memory is implemented in two ways, first by adding extra dice at fixed $M$, second by increasing $M$.

\subsection{Game A: Adding dice}
The face value in this game is determined by rolling multiple dice and choosing the one with the maximum face value. One could imagine independent ecosystems co-existing, with the final `reset' only occurring when the most stable subsystem collapses. 
\begin{figure}[H]
         \centering
         \includegraphics[trim={0cm 5.9cm 0cm 6cm},clip,width=\textwidth]{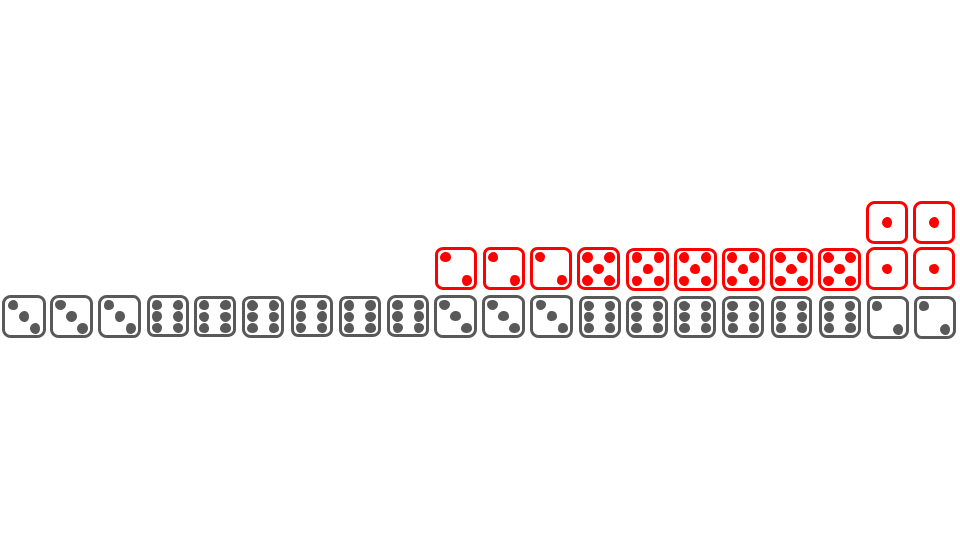}
         \caption{One possible unfolding of SSM game A with $T=20$ and $M=6$. At $t=3$ we roll 6 which adds an extra die. At $t=13$ we roll six again which adds a third die. The bottom row is the observable, the other rows show dice rolls which are not observed.}
         \label{fig:ssma_example}
\end{figure}

In the limit of large $M$, the expected number of dice at time $t$ is
\begin{equation}\label{eqn:numdice}
r(t) = A \exp\left( \frac{t}{M^2}\right)
\end{equation}
where $A = e^{1-\gamma} \sim 1.5$. The number of dice grows exponentially, with doubling time $\sim M^2 \ln2$. The more faces a die has, the longer we have to wait to land on a specific one. The expected face value at time $t$ is
\begin{equation}\label{eqn:ssmface}
E[f:M\gg1](t) = M \frac{A\exp\left( \frac{t}{M^2}\right)}{1 + A\exp\left( \frac{t}{M^2}\right)}
\end{equation}
a sigmoid function of $t/M^2$. At large values of $t$ the expectation tends to $M$, we have so many dice we are virtually guaranteed to roll at least one $M$. 

\begin{figure}[H]
         \centering
         \includegraphics[width=\textwidth]{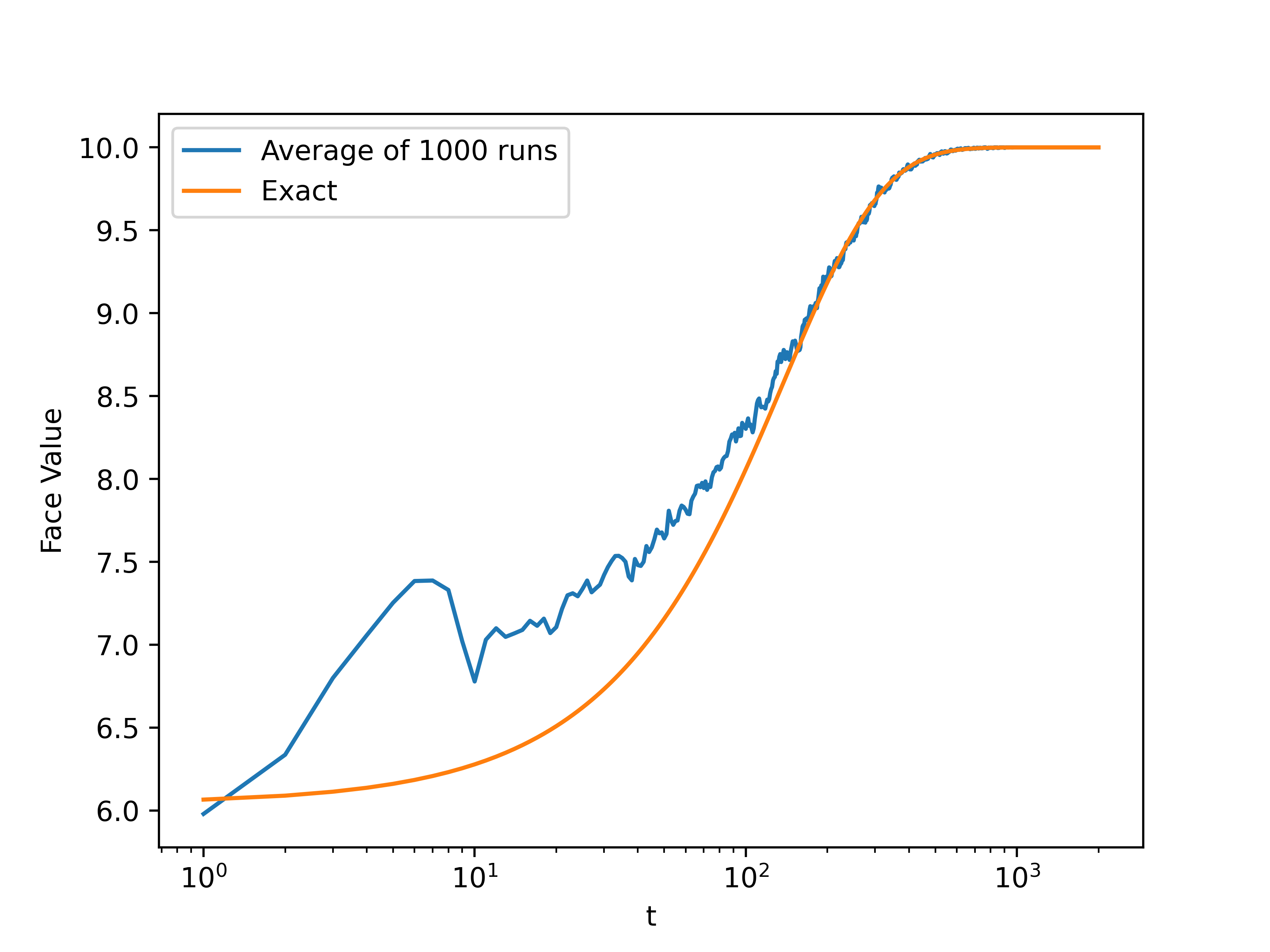}
         \caption{Average face value in the SSM game as a function of $t$ for an $M=10$ sided dice, averaged over $N=1000$ independent instances.}
         \label{fig:ssm_run}
\end{figure}
Figure \ref{fig:ssm_run} shows the results of 1000 simulations of the game with $M = 10$ compared to the `exact' answer, equation \ref{eqn:ssmface}. We observe convergence to the upper bound $M$ at a rate that is roughly linear in log time. Despite the exponential growth in the number of dice suggested by equation \ref{eqn:numdice}, the expected face value grows much more slowly. Such slow convergence is seen in more complex evolutionary models, especially the Tangled Nature Model \citep{christensen2002tangled, arthur2017tangled} and its variants \citep{arthur2017entropic,arthur2022selection}. There, it arises from the simulated ecosystem successively crossing `entropic barriers' \citep{becker2014evolution}. Each time a barrier is crossed the system is likely to land in a more stable configuration  with a higher barrier. This behaviour is also observed in physical systems like spin glasses, colloids and high temperature superconductors \citep{sibani2021record}. 

This model shows that there is competition between the growth in the number of dice against growth in time taken between trials. What it  suggests is that selection plus accumulation leads to slow growth in stability. In contrast to SBS and SS, SSM implies that older inhabited planets should be more habitable, so our presence on Earth is not an observer effect but a statistically more likely outcome. 

\subsection{Game B: Increasing M}

Similar results are observed with other types of `memory' . Here, instead of adding extra dice over time we have just one die and add  extra faces to it. A rough analogy to a real ecosystem is to assume that species diversity is not lost after each collapse (dice roll) and that species persist at low abundance, in dormant states or isolated refugia. Reaching a fitness peak (hitting the max value of $M$) generates more diversity and allows ecosystems to explore more of the so-called fitness landscape \citep{arthur2017decision}. Thus each reset has the potential to find a more stable state because the space of possibilities is wider. 
\begin{figure}[H]
         \centering
         \includegraphics[trim={0cm 7.5cm 0cm 7.5cm},clip,width=\textwidth]{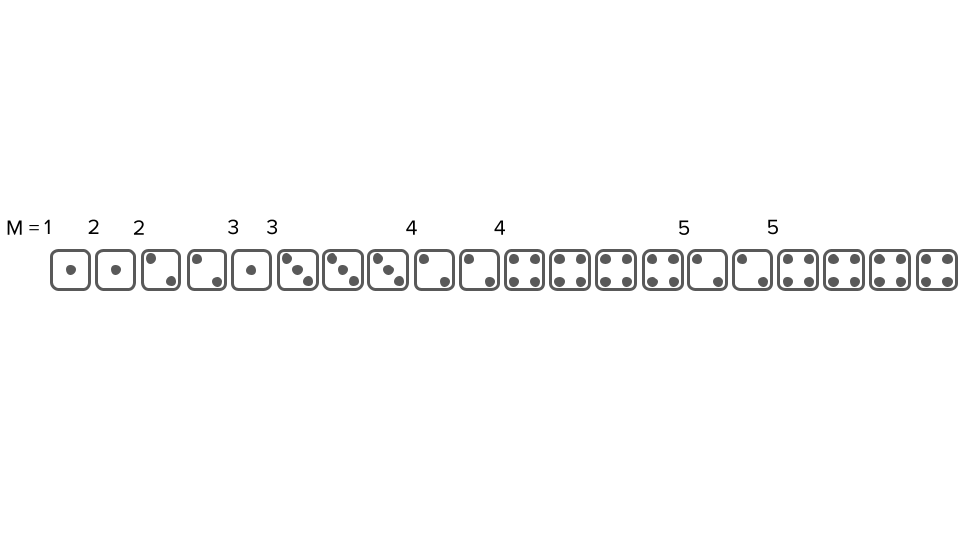}
         \caption{One possible unfolding of SSM game B with $T=20$. The bottom row shows the actual face values and the top row shows the number of sides of the die. For example at $t=11$ we roll a 4 on a 4-sided die, increasing the number of sides to 5 for the next roll.}
         \label{fig:ssmb_example}
\end{figure}

\begin{figure}[H]
         \centering
         \includegraphics[width=\textwidth]{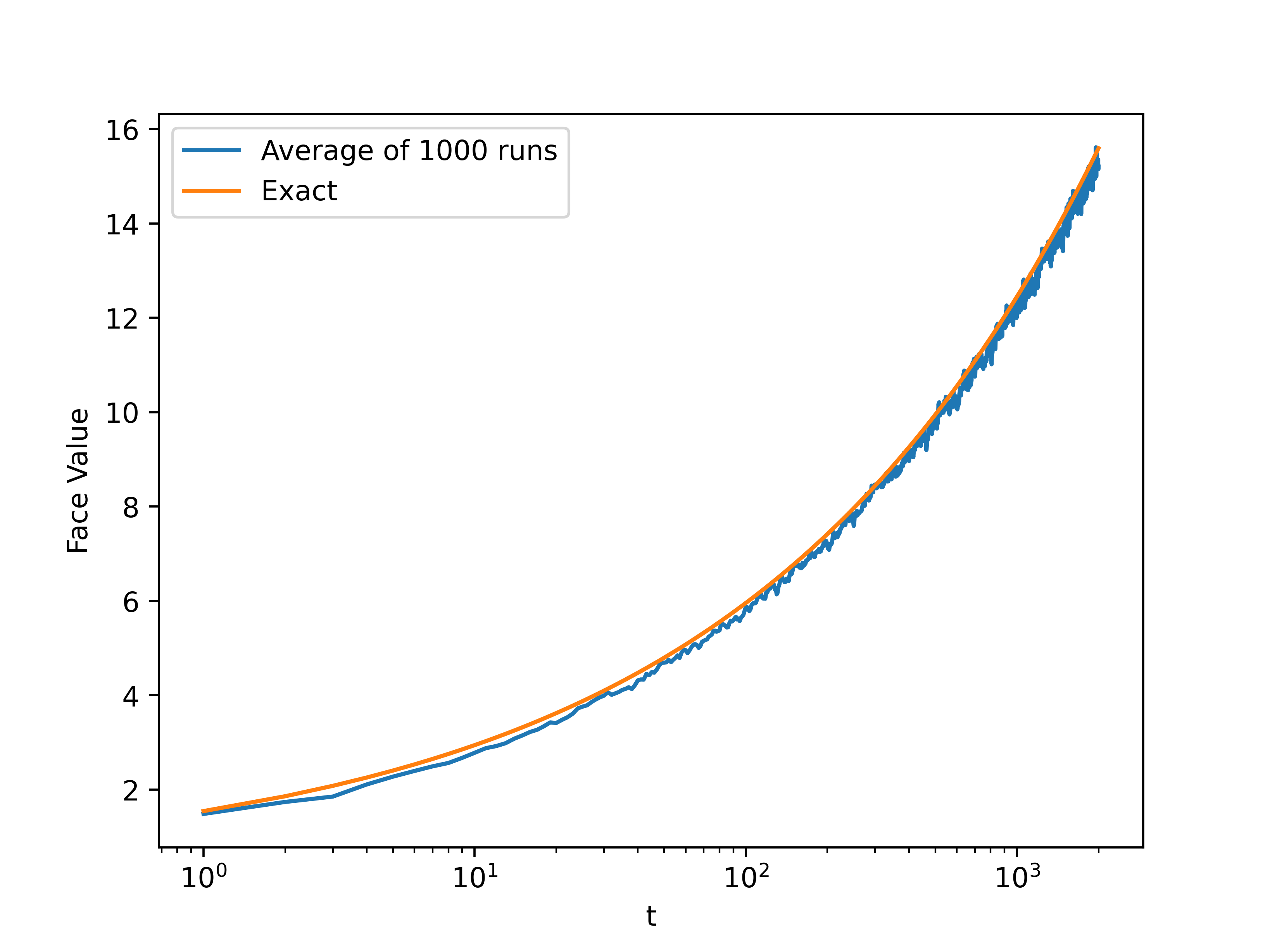}
         \caption{Average face value in SSM game B as a function of $t$, averaged over $N=1000$ independent instances.}
         \label{fig:ssm2_run}
\end{figure}
At time $t$ the expected face value is
\begin{equation}\label{eqn:ssmbface}
E[f](t) = \frac{2\sqrt[3]{6t} + 1}{3}
\end{equation}
Figure \ref{fig:ssm2_run} shows the results of simulations of the game compared to the exact answer, equation \ref{eqn:ssmbface}. Unlike Game A there is no convergence and the expected face value grows without bound, though fairly slowly. Again there is a trade off between increasing $M$ by performing a large number of trials and the time it takes to complete those trials, though in this version there is no limiting value of fitness. This is again reminiscent of Tangled Nature Model dynamics \citep{becker2014evolution,arthur2017entropic,arthur2022selection} and other physical systems which cross energetic or entropic barriers \citep{sibani2021record}. Again the implication is that older systems are more likely to be stable with higher habitability.

\section{Conclusion}
These three mechanisms give reasonable ideas about what to expect when surveying large catalogues of inhabited planets or looking at a single inhabited planet at a random point in its history. The first two (SBS, SS) have no role for life in enhancing stability, implying that the stability properties of inhabited planets are down to observer effects. The third mechanism (SSM) suggests that inhabited planets have properties conducive to stability \emph{because of their history of inhabitance}. 

This idea has appeared previously as `The Inhabitance Paradox' in \cite{goldblatt2016inhabitance} and is closely related to the idea of the Gaian bottleneck \citep{chopra2016case}. The Gaian bottleneck suggests that life must `seize the reins' and exert a stabilising effect early in a planet's history or go extinct due to deteriorating geophysical conditions. The Inhabitance Paradox says that for a planet to be habitable, it must be inhabited, suggesting that life must not only grab, but also keep the reins. Echoing older work \citep{lovelock1989geophysiology}, planetary habitability may be all or nothing. 

The SSM games suggests that combining the sequential selection algorithm with some method of storing diversity will tend to generate more stable systems. We have argued previously in \cite{arthur2022selection} that such diversity stores are widespread on Earth, for example: microbial seed banks, dormancy \citep{lennon2011microbial} and lateral gene transfer \citep{goldenfeld2007biology} all contribute to the maintenance of microbial diversity. It is not a huge leap to imagine similar processes on exoplanets. We propose that Gaia - the stabilising and symbiotic feedback of life and the environment - can be born through this kind of natural, but non-Darwinian, selection. 

The implication of all selection principles for exoplanet surveys is that older planets are more likely to host the large and influential biosphere necessary to produce a detectable biosignature. SSM makes the additional prediction that, on average, habitability on an inhabited planet improves over time. Given that this is achieved through life's influence on the environment, this suggests biosignatures get stronger over time. SSM, along with the Gaian bottleneck and Inhabitance paradox, suggest that `goldfish bowl' planets \citep{wilkinson2003fundamental} are rare, planets either have abundant life or are uninhabited. These are statistical tendencies and we can have cases where Gaia is dormant and life is clinging on but could later rebound to become abundant, similar to the rebound after the Snowball Earth periods \citep{hoffman1998neoproterozoic}. Thus  failure to detect a biosignature does not mean the absence of life (an absence is impossible to prove in any case). However, SSM does suggest that periods of global dormancy are short (Snowball Earths lasted $\sim$10My) and an inhabited planet is likely to be observed with an abundant biosphere, and therefore a large biosignature.

Assuming that a large catalogue of planets, their ages and inhabitance status was available, SSM could be tested by examining the strength of biosignatures and the rate of inhabitance. SSM would predict a constant rate of inhabitance over time and that the average `strength' of biosignatures \citep{seager2013biomass, nicholson2022predicting} slowly increases. This is only a first order approximation, which is complicated by a number of factors from stellar evolution to runaway greenhouse effects, see e.g. \citep{Arthur:2023} for a discussion on how these ideas interact with temperature regulation and location within the habitable zone. Generally, SSM is optimistic for life detection, suggesting that inhabitance is associated with strong biosignatures and so if life gets through the initial bottleneck it will persist while producing a large signature.

\section*{Author Contributions}
Both authors contributed equally to this manuscript.

\section*{Conflict of Interest}
No conflicts of interest are declared

\section*{Funding}
This work was supported by a Leverhulme Trust research project grant [RPG-2020-82].

\appendix
\section*{SBS Game}
At $t=1$ all of the dice are in play and the average face value (over very large $N$ or many different realisations of the game) is 
\begin{equation*}
(1+2+\ldots+M)/M.
\end{equation*}
At $t=2$ all of the dice showing 1 on the top face are removed. Since our survey is of inhabited planets, the average face value is now
\begin{equation*}
(2+3+\ldots+M)/(M-1)
\end{equation*}
Following the pattern, at time $t \leq M$ the expected face value is
\begin{equation}
\frac{(t + (t+1) + \ldots + M)}{(M-t)} = \frac{M+t}{2}
\end{equation}
So the average face value increases linearly with time.

\section*{SS Game}

 In the Sequential Selection game, the chance of the die showing $k$ is  proportional to the probability of rolling a $k$, $p(k)$, times the number of `slots' where the observation could occur e.g. if the die is showing 3 this could be an observation of the die on the first, second or third step where it is face up. We normalise this probability and compute the expected value for the top face as 
\begin{equation}\label{eqn:ss_eqn}
\sum_{k=1}^M k \frac{k p(k)}{\sum_{k=1}^M k p(k) } = \frac{2M+1}{3}
\end{equation}
Where we use, $p(k) = 1/M$. Note this is larger than the expected value of a single dice roll, $\frac{M+1}{2}$ for $M>1$.

\section*{SSM Game A}

First we need is the probability to get the face value $f$ when rolling $r$ dice and applying the rule $f = \max(x_1, x_2, \ldots, x_r)$. This is
\begin{equation*}
p_r(f) = \sum_{i=1}^r \binom{r}{i} p(f)^i p(x<f)^{r-i}
\end{equation*}
where $p(f) = 1/M$ and $p(x<f) = \frac{f-1}{M}$. This is just the probability to get at least one $f$ and nothing higher, multiplied by a combinatoric factor. To simplify this, consider arranging all the possible outcomes of $r$ rolls in an r-dimensional hypercube. The number of ways to obtain $f$ is given by the difference in volumes between an $f$ and $f-1$ sided hypercube so
\begin{equation*}
p_r(f) = \frac{f^r - (f-1)^r}{M^r}
\end{equation*}
The expectation for the face value is therefore
\begin{equation*}
E[f] = \sum_{k=1}^M k \frac{k^r - (k-1)^r}{M^r}
\end{equation*}
Writing out the sum explicitly
\begin{align*}
&1.1^r + 2.2^r + 3.3^r + \ldots + M.M^r \\ \nonumber
-(&1.0^r + 2.1^r + 3.2^r + \ldots + M.(M-1)^r)
\end{align*}
Shows that we can regroup and rewrite as
\begin{equation}\label{eqn:summation}
E[f] = \frac{1}{M^r} \left( M^{r+1} - \sum_{k=1}^{M-1} k^r\right) 
\end{equation}
The sum can be simplified using Faulhaber's formula \citep{weisstein}
\begin{equation*}
\sum_{k=1}^{M-1} k^r = \frac{(M-1)^{r+1}}{r+1} + \frac{(M-1)^r}{2} + O(M^{r-1})
\end{equation*}
where the lower order terms are complicated coefficients involving the Bernoulli numbers. Substituting and taking the limit of large $M$ we get
\begin{equation}
E[f|r,M\gg1] = M\frac{r}{r+1} + \frac{1}{2}
\end{equation}
This result is of some interest to recreational mathematicians \citep{youtubedice}.

In a game with $r$ dice, the expected number of dice rolls before hitting the value $M$, where we add an extra die, is $1/p_r(M)$. Therefore, the expected time spent playing with exactly $r$ dice is 
\begin{equation*}
T(r) = \frac{ \sum_{i=1}^M i p_r(i)}{p_r(M)}. 
\end{equation*}
For large $M$ (using equation \ref{eqn:summation}) this is
\begin{equation*}
T(r) \simeq \frac{M^2}{r+1}.
\end{equation*}
To calculate the expected face value at $t$, we first compute the expected number of dice at time $t$ by solving
\begin{equation*}
\sum_k^{r} T(k) = t
\end{equation*}
for $r$. Using the large $M$ approximation
\begin{equation*}
\sum_k^{r} T(k) \simeq M^2 \sum_k^{r} \frac{1}{i+1} = M^2 (H_r - 1)
\end{equation*}
where $H_r$ is the $r^{th}$ harmonic number. This has a standard approximation, valid for large $r$ but quite accurate even at $r=1$: $H_r \simeq \gamma + \ln(r)$, where $\gamma$ is the Euler-Mascheroni constant. Substituting and solving for $r$ gives
\begin{equation*}
r(t) = \exp\left( \frac{t}{M^2} + 1 - \gamma\right) = A \exp\left( \frac{t}{M^2}\right)
\end{equation*}
where $A = \exp(1 - \gamma)$. The expected face value at time $t$ is the expected face value with $r(t)$ dice. Still working in the large $M$ limit this is
\begin{equation*}
E[f:M\gg1](t) \simeq M\frac{r}{r+1} = M \frac{A\exp\left( \frac{t}{M^2}\right)}{1 + A\exp\left( \frac{t}{M^2}\right)}
\end{equation*}

\section*{SSM Game B}
For an $M$-sided die, the expected number of rolls required to hit the $M$ face is  $M$. The expected time for a face to be up is $\frac{M+1}{2}$, so the expected time spent playing with an $M$-sided die is
\begin{equation}
T(M) = M\frac{M+1}{2}
\end{equation}
Summing up the wait times from each $M$ gives the total duration of the experiment
\begin{equation}
t = \sum_{i=1}^M \frac{i(i+1)}{2} = \frac{1}{2} \left( \frac{M(M+1)(2M+1)}{6} + \frac{M(M+1)}{2} \right)
\end{equation}
Keeping only the terms of highest order in $M$ and solving for $t$ gives
\begin{equation}
M = \sqrt[3]{6t}
\end{equation}
Substituting into equation \ref{eqn:ss_eqn} gives
\begin{equation}
E[f](t) = \frac{2\sqrt[3]{6t} + 1}{3}
\end{equation}

\section*{All Sequential Games}
\begin{figure}[H]
         \centering
         \includegraphics[width=\textwidth]{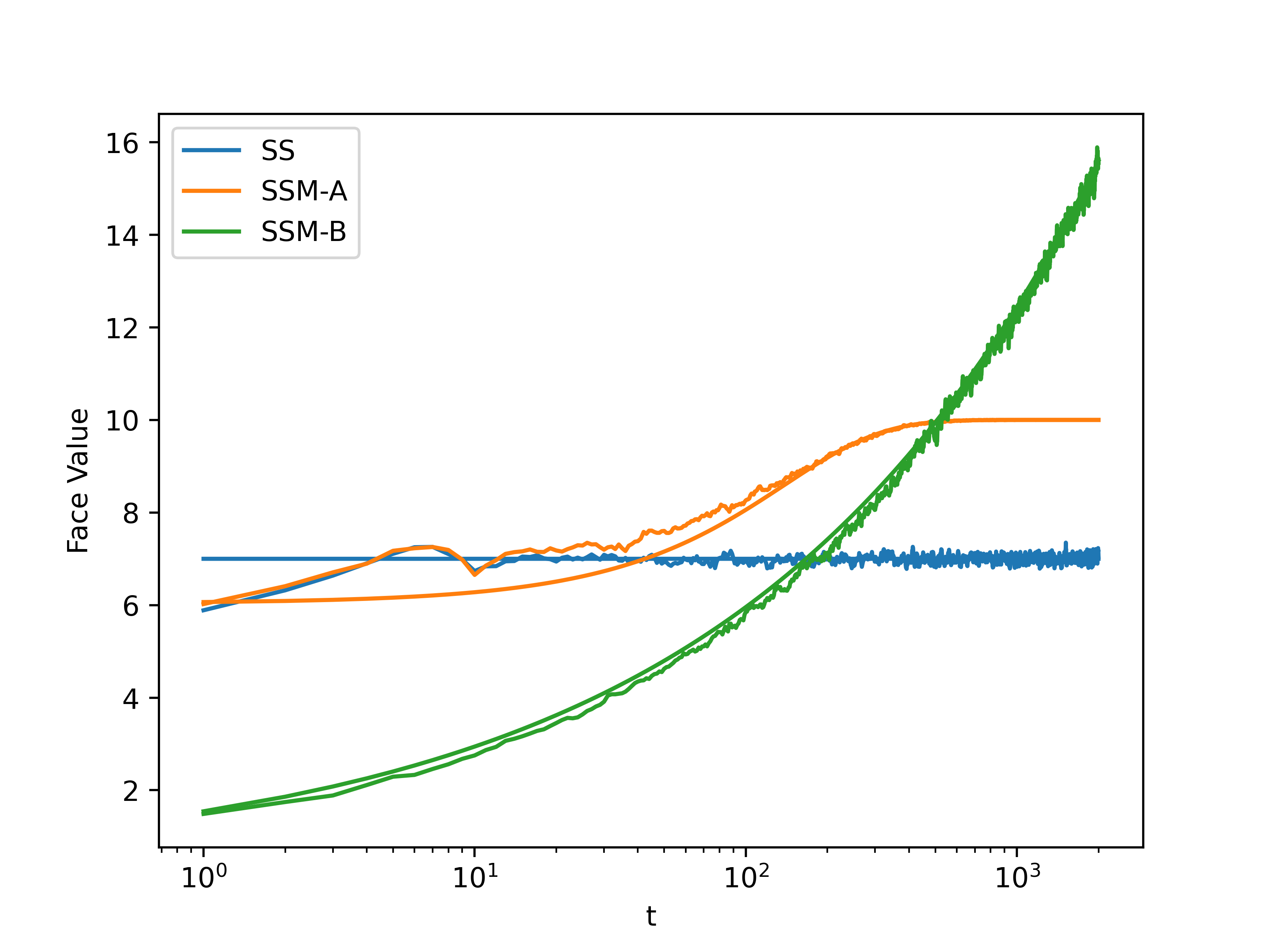}
         \caption{Average face value in each of the sequential games  as a function of $t$, averaged over $N=1000$ independent instances. Combines Figures 4, 6 and 8 from the main text.  The SBS game without modification would appear as a straight line that ends after $M$ time steps. }
         \label{fig:ssm2_run}
\end{figure}

\bibliographystyle{plainnat}
\bibliography{arxiv}

\end{document}